\documentclass[12pt]{article}
\usepackage{epsfig,epsf,xcolor,verbatim}

\begin{document}

\begin{center}
{\Large\sc Travelling times in  scattering by obstacles}
\end{center}

\begin{center}
{\sc Lyle Noakes and Luchezar Stoyanov}
\end{center}

\newcommand{\R}{{\sf I\hspace{-.15em}R}}
\def\C{{\bf C}}
\def\e{\emptyset}
\def\ds{\partial S}
\def\cq{\overline{Q}}
\def\sn{{\bf S}^{n-1}}
\def\hg{\Gamma}
\def\ssn{{\sn}\times {\sn}}
\def\do{\partial \Omega}
\def\dk{\partial K}
\def\dl{\partial L}
\def\tts{\tilde{S}}
\def\ttp{\tilde{p}}
\def\tu{\tilde{U}}
\def\hs{\hat{S}}
\def\hp{\hat{p}}
\def\dr{\frac{\partial r}{\partial x_1}}
\def\ll{{\cal L}}
\def\hm{\hat{M}}
\def\pp{{\cal P}}
\def\tt{{\cal T}}
\def\ds{\partial S}
\def\ss{{\cal S}}
\def\vv{{\cal V}}
\def\aa{{\cal A}}
\def\bb{{\cal B}}
\def\nn{{\cal N}}
\def\dd{{\cal D}}
\def\ee{{\cal E}}
\def\uu{{\cal U}}
\def\rr{{\cal R}}
\def\kk{{\cal K}}
\def\cc{C_0^{\infty}}
\def\ot{(\omega,\theta)}
\def\oto{(\omega_0,\theta_0)}
\def\toto{(\tilde{\omega}_0, \tilde{\theta}_0)}
\def\ts{\tilde{\sigma}_0}
\def\tx{\tilde{x}}
\def\txi{\tilde{\xi}}
\def\got{\gamma(\omega,\theta)}
\def\ggot{\gamma'(\omega,\theta)}
\def\omo{(\omega,\omega)}
\def\du{\partial u_i}
\def\dfu{\frac{\partial f}{\partial u_i}}
\def\dou{\frac{\partial \omega}{\partial u_i}}
\def\dttu{\frac{\partial \theta}{\partial u_i}}
\def\ooo{(\omega_0,-\omega_0)}
\def\otp{(\omega',\theta')}
\def\oo{{\cal O}}
\def\pr{{\rm pr}}
\def\cK{\hat{K}}
\def\cL{\hat{L}}
\def\ff{{\cal F}}
\def\fk{{\cal F}^{(K)}}
\def\fl{{\cal F}^{(L)}}
\def\kkr{{\cal K}^{\mbox{reg}}}
\def\kkro{{\cal K}_0^{\mbox{reg}}}
\def\trapk{\mbox{\rm Trap}(\Omega_K)}
\def\trapl{\mbox{\rm Trap}(\Omega_L)}
\def\sock{S^*(\stackrel{\circ}{\Omega}_{\cK})}
\def\oock{\stackrel{\circ}{\Omega}_{\cK}}
\def\socl{S^*(\stackrel{\circ}{\Omega}_{\cL})}
\def\oocl{\stackrel{\circ}{\Omega}_{\cL}}
\def\sok{S_b^*(\Omega_K)\setminus \trapk}
\def\sol{S_b^*(\Omega_L)\setminus \trapl}
\def\sbok{S_b^*(\Omega_K)}
\def\sbol{S_b^*(\Omega_L)}
\def\hr{\hat{\rho}}
\def\G{{\cal G}}
\def\tg{\tilde{\gamma}}
\def\Vo{\mbox{Vol}}
\def\dkso{\partial K^{(\infty)}}
\def\dlso{\partial L^{(\infty)}}
\def\gg{{\cal g}}
\def\sl{{\cal SL}}
\def\tkn{\mbox{\rm Trap}^{(n)}(\dk)}
\def\tln{\mbox{\rm Trap}^{(n)}(\dl)}
\def\wn{{\cal WN}}
\def\i{{\bf i}}
\def\te{{\cal T}^{(ext)}}
\def\ndk{{\cal N}_b^*(\dk)}
\def\ndl{{\cal N}_b^*(\dl)}
\def\dr{\frac{\partial r}{\partial z_1}}
\def\endofproof{{\rule{6pt}{6pt}}}
\def\Box{\endofproof}
\def\su{S^*(\R^n\setminus U)}
\def\gk{\gamma_K}
\def\gl{\gamma_L}
\def\la{\left\langle}
\def\ra{\right\rangle}
\def\kfin{\kk_0^{({\mbox{\footnotesize\rm fin}})}} 
\def\dt{\dot{T}}
\def\ep{\epsilon}
\def\kfi{\kk^{({\mbox{\footnotesize\rm fin}})}} 
\def\stk{\Sigma_3^{(K)}}
\def\stl{\Sigma_3^{(L)}}
\def\SU{S^*(\R^n\setminus U)}
\def\tkm{\tt_k^{(m)}}
\def\ukm{U_k^{(m)}}
\def\Pkm{\Psi_k^{(m)}}
\def\nkm{N_k^{(m)}}
\def\di{\displaystyle}
\def\nk{N^{(K)}}
\def\nl{N^{(L)}}
\def\mk{M^{(K)}}
\def\ml{M^{(L)}}
\def\ep{\epsilon}
\def\Gk{G^k}
\def\uk{U^{(K)}}
\def\dist{\mbox{\rm dist}}
\def\diam{\mbox{\rm diam}}
\def\con{\mbox{\rm const}}
\def\bs{\bigskip}
\def\ms{\medskip}
\def\SK{S^*_{K}(S_0)}
\def\SL{S^*_{L}(S_0)}

\def\te{\tilde{e}}
\def\tx{\tilde{x}}
\def\ty{\tilde{y}}
\def\grad{\; \mbox{\rm grad} \,}
\def\spp{\vspace{5pt}\noindent}
\newtheorem{definition}{Definition}
\newtheorem{construction}{Construction}
\newtheorem{example}{Example}
\newtheorem{lemma}{Lemma}
\newtheorem{theorem}{Theorem}
\newtheorem{corollary}{Corollary}
\newtheorem{proposition}{Proposition}

\noindent
{\sc Abstract.} 

\medskip

\footnotesize
\noindent
The paper deals with some problems related to recovering information about an obstacle in an Euclidean space from certain 
measurements of lengths of generalized geodesics in the exterior of the obstacle. The main result is that if two obstacles satisfy some generic
regularity conditions and have (almost) the same travelling times, then the generalized geodesic flows in their exteriors are conjugate 
on the non-trapping part of their phase spaces with a time preserving conjugacy. Apart from that a constructive algorithm is described that
shows how to recover an obstacle consisting of two convex domains in the plane from traveling times.

\medskip



\normalsize

\section{Introduction}\label{sec1}
\renewcommand{\theequation}{\arabic{section}.\arabic{equation}}

Let $K$ be a compact subset of ${\R}^n$ whose boundary $\partial K$ is a $C^\infty$ manifold of dimension $n-1$.  
Suppose that  $\Omega_K = \overline{{\R}^n\setminus K}$ is connected. Let $S_0$ be a large sphere\footnote{Our results extend 
easily to the case where $S_0$ is the boundary of some other $C^\infty$ strictly convex subset of $\R ^n$ containing $K$ in its interior.} 
in $\R^n$ bounding  an open ball that contains $K$. For any $x,y\in S_0$, an $(x,y)$-geodesic in $\Omega _K$ 
is a generalised geodesic (in the sense of Melrose and  Sj\"ostrand \cite{kn:MS1}, \cite{kn:MS2}) from $x$ to $y$. 
The length of the part of $\gamma $ from $x$ to $y$ is denoted by $t_\gamma$.

This article deals with some  problems related to recovering information about the obstacle $K$ from certain
measurements of lengths of generalised geodesics in the exterior of $K$. These problems have similarities with 
various problems on metric rigidity in Riemannian geometry -- see \cite{kn:SU}, \cite{kn:SUV} and the 
references there for more information.

\spp

More specifically, we consider the problem of obtaining information about $K$ from travelling times of $(x,y)$-geodesics, especially from the 
{\it travelling times spectrum} $\tt_K$ of $K$, defined as follows. 

\begin{definition} Let ${\cal T}_K\subset S_0\times S_0\times [0,\infty )$ be the  set all triples $(x,y,t_\gamma )$ where $\gamma $ is an $(x,y)$-geodesic. \endofproof
\end{definition}
%

\spp
For any $x,y\in S_0$, define ${\cal T}_K(x,y):=\{ t\in [0,\infty ):(x,y,t)\in {\cal T}_K\} $. 
Then ${\cal T}_K(x,y)={\cal T}_K(y,x)$ for all $(x,y)\in S_0\times S_0$.

\textcolor{black}
The {convex hull}  $\widehat{K}$ of $K$ is easily found from ${\cal T}_K$ as follows. 
\begin{example}\label{ex2} Given $x_0\in S_0$ let $\nu _{x_0}$ be the inward-pointing unit normal to $S_0$ at $x_0$.  
Given $r\geq 0$, let $H_r$ the the hyperplane through $x_0+r\nu _{x_0}$ and orthogonal to $\nu _{x_0}$. 
Then $H_0$ has no common points with the interior of $ \widehat K$. 
Let $r(x_0)=\sup \{ r:  H_s\cap K=\emptyset \hbox{~for all~}s\leq r\}$. 
Then, as $x_0$ varies over $S_0$, the $\partial H_{r(x_0)}$ are precisely the supporting hyperplanes of $\widehat K$. 
But $H_s\cap K=\emptyset \Longleftrightarrow \Vert x-y\Vert \in {\cal T}_K(x,y)$ for all $x,y\in S_0\cap H_s$, and so the 
supporting hyperplanes are characterised in terms of ${\cal T}_K$.  \endofproof \end{example} 

\spp
More generally, given ${\cal T}_K$ a hyperplane 
$H$ in $\R ^n$ is said to be {\em vacuous} when, for all intervals $xy\subset H$ with $x,y\in S_0$,  
$\Vert x-y\Vert \in {\cal T}_K(x,y)$. (When $H\cap S_0=\emptyset$, $H$ is said to be 
{\em trivially vacuous}.) The supporting hyperplanes of Example \ref{ex2} are a special kind 
of vacuous hyperplane.  
The set $V$ of all vacuous hyperplanes is an open subset of real projective $n-1$-space, 
and $\partial V$ is the space of hyperplanes $H$ whose intersections with $K$ are nonempty with  
$H$ tangent to $K$. So ${\cal T}_K$ determines the envelope $Y\subseteq \partial K$ of $\partial V$.  

\spp
Even when it is easy to construct $K$, it is usually difficult to say which set-valued functions have the form ${\cal T}_K$. 
\begin{example}\label{ex1} Let $K$ be connected and strictly convex.  
Because $K$ is strictly convex, for any $x\in S_0$ the nearest point $z_x$ in $K$ to $x$ is uniquely defined and 
$C^\infty$ as a function of $x$. Notice that the $C^\infty$ function $f:S_0\rightarrow \R$ given by $f(x)=2\Vert x-z_x\Vert $ 
is characterised in terms of ${\cal T}_K$ by  
$f(x)=\min {\cal T}_K(x,x)$.  

\spp
For $v$ tangent to $S_0$ at $x$ we have $df_x(v)=2\langle v, u_x\rangle $, where $u_x$ is the unit vector from $z_x$ to $x$. 
This determines $u_x$ and then $z_x$ in terms of $f$. Because any $z\in \partial K$ has the form $z_x$ for some $x\in S_0$, 
it follows that $\partial K$ is recovered  from the  ${\cal T}_K(x,x)$. But $K$ can also be found as $\widehat K$  from ${\cal T}_K$ 
as in Example \ref{ex2}, namely by testing the condition $\Vert x-y\Vert \in {\cal T}_K(x,y)$. 

\spp
Since the ${\cal T}_K(x,x)$ and ${\cal T}_K(x,y)$ are independently sufficient to reconstruct the obstacle $K$, there is an implicit constraint 
on ${\cal T}_K$. Indeed this understates the complexity of the conditions that ${\cal T}_K$ needs to satisfy: for this especially simple case it 
can easily be seen that there are uncountably many different algorithms for reconstructing $K$ from the ${\cal T}_K(x,y)$. 
\endofproof
\end{example}

\spp
When the set-valued function ${\cal T}_K$ is known 
only approximately, different algorithms for recovering $K$ (such as in Example \ref{ex1}) usually give different 
estimates of $K$. Better estimates could be made if we understood the internal structure of ${\cal T}_K$.

If $K$ and $L$ are two obstacles contained in the interior of $S_0$, we will say that $K$ and $L$  {\it have almost the same travelling times} if 
$\tt_K(x,y) = \tt_L(x,y)$ for almost all  $(x,y) \in S_0\times S_0$ (with respect to the Lebesgue measure on $S_0\times S_0$).
Our main result  shows that if $K$ and $L$ satisfy some mild  non-degeneracy conditions and have almost the same travelling times, then the 
generalised geodesic flows in their exteriors are conjugate on the non-trapping part of their phase spaces, with a time preserving conjugacy. 

A different inverse scattering problem was studied  in  \cite{kn:St3} and \cite{kn:St2}. The observables used there to get geometric information
about an obstacle $K$ were the so called sojourn times of $\ot$-rays $\gamma$ for pairs $(\omega, \theta)$ of unit vectors in $\R^n$.
By an $\ot$-ray in the exterior $\Omega_K$ of $K$ we mean a billiard trajectory in $\Omega_K$ incoming from infinity with direction $\omega$
and outgoing to infinity with direction $\theta$. If  $\Pi_{\omega}$ is the hyperplane in $\R^n$ tangent to $S_0$ and orthogonal to $\omega$,  denote by
$H_{\omega}$ the open half-space  determined by  $\Pi_{\omega}$ and having $\omega$ as an inner normal. 
For an $\ot$-ray $\gamma$ in $\Omega_K$, the {\it sojourn time} $T_{\gamma}$ is $T'_{\gamma} - 2a$, where $T'_\gamma$ is the length of the part of 
$\gamma$ contained in  $H_{\omega}\cap H_{-\theta}$ and $a$ is the radius of the sphere $S_0$.

Denoting by $\sl_K\ot$ the set of sojourn times $T_\gamma$ of all $\ot$-rays $\gamma$ in $\Omega_K$, the family
 $\sl_K =  \{ \sl_K\ot\}_{\ot}$, where $\ot$ runs over $\ssn$, is called the {\it scattering length spectrum} of $K$.
It was proved in \cite{kn:St3} that if two obstacles $K, L$  in $\R^n$, with sufficiently `regular boundaries', have almost the 
same scattering length spectrum, then their generalized geodesic flows are conjugate on the non-trapping parts 
of their phase spaces. This result was then used to derive
 some properties of obstacles that can be  recovered from the scattering length spectrum $\sl_K$  -- see Corollary \ref{cor1} below for some details.

In general  the scattering length spectrum does not determine $K$ -- see the example of M. Livshits 
in Ch. 5 of \cite{kn:M}. The same applies to the travelling times spectrum $\tt_K$.

Let $\fk_t : \dt_b^*(\Omega_K) = T_b^*(\Omega_K)\setminus \{ 0\} \longrightarrow \dt_b^*(\Omega_K)$ be the {\it generalized geodesic flow}
generated by the principal symbol of the wave operator in $\R\times \Omega_K$ (see \cite{kn:MS1}, \cite{kn:MS2}, or \S 24.3 in \cite{kn:H}),
where $T_b^*(\Omega_K) = T^*(\Omega_K)/\sim$ is the  quotient space with respect to the equivalence relation: $(x,\xi) \sim (y,\eta)$
iff $x = y$ and either $\xi = \eta$ or $\xi$ and $\eta$ are symmetric with respect to the tangent plane to $\dk$ at $x$. Denote by
$S_b^*(\Omega_K)$ the image of the {\it unit cosphere bundle}  $S^*(\Omega_K)$. In what follows we identify $T^*(\Omega_K)$ and 
$S^*(\Omega_K)$ with their images in $T_b^*(\Omega_K)$. For any $(y,\eta) \in T^*(\Omega_K)$ set $\mbox{pr}_1 (y,\eta) = y$ and $\mbox{pr}_2 (y,\eta) = \eta$.

Here we make the following assumption about $K$: for each $(x,\xi)\in \dt^* (\partial K)$ if the curvature of $\partial K$ at $x$ vanishes of infinite order 
in direction $\xi$, then all points $(y,\eta)$ sufficiently close to $(x,\xi)$ are diffractive points (roughly speaking, this means that $\dk$ is convex at $y$ 
in the direction of $\eta$). Let $\kk$ be the {\it class of obstacles with this property}.  For $K\in \kk$ the flow $\fk_t$ is well-defined and continuous (\cite{kn:MS2}).

A point $\sigma = (x,\omega)\in S_b^*(\Omega_K)$ is called {\it non-trapped} if both $\{ \mbox{pr}_1(\fk_t(\sigma)) : t\leq 0\}$ and 
$\{ \mbox{pr}_1(\fk_t(\sigma)) : t\geq 0\}$ in $\Omega_K$ are unbounded curves in $\R^n$.  Let $\trapk$ be the {\it  set  of all trapped points}. 
In general $\trapk$ may have positive Lebesgue measure and non-empty interior  
in $S_b^*(\Omega_K)$ (see e.g. Livshits' example mentioned above).   $K$ is called a {\it non-trapping} obstacle if  $\trapk = \e$.

Set $S^*_+(S_0) = \{ (x,u) : x\in S_0, u\in \sn \;, \; \la x,u\ra < 0 \}$, and let 
$S^*_{K}(S_0)$ be {\it the set of all points of $S^*_+(S_0)$ which define non-trapped geodesics} in $\Omega_K$.
Consider the cross-sectional map  $\pp_K :  \SK \longrightarrow S^*(S_0)$ defined by the shift along the flow $\fk_t$.
Given  $\sigma = (x,\xi)\in S_b^*(\Omega_K)$, set 
$\gamma_K(\sigma) = \{ \mbox{pr}_1(\fk_t(\sigma)) : t\in \R\} $.
A trajectory of this kind is called a {\it simply reflecting ray} if it has no tangencies to $\dk$.

Let $\gamma$ be a  $(x_0,y_0)$-geodesic in $\Omega_K$ for some $x_0,y_0 \in S_0$, which is a simply reflecting ray. Let $\omega_0\in \sn$ be the 
(incoming) direction of $\gamma$ at $x_0$.  We will say that $\gamma$ is {\it regular} if the differential of map 
$\sn \ni \omega \mapsto \pr_1(\pp_K (x_0,\omega))\in S_0$ is a submersion at $\omega = \omega_0$, i.e. its differential at that point  has rank $n-1$.

Denote by $\ll_0$ the { class of all obstacles} $K\in\kk$  such that $\dk$ does not contain non-trivial open flat subsets and $\gamma_K(x,u)$ is a regular simply 
reflecting ray for almost all $(x,u) \in S^*_+(S_0)$ such that $\gamma(x,u)\cap \dk\neq \e$. One can derive from Ch. 3 in \cite{kn:PS1} that  $\ll_0$ is of 
second Baire category  in $\kk$ with respect to the $C^\infty$ Whitney topology in $\kk$. That is, generic obstacles $K\in \kk$ belong to the class $\ll_0$. 

The main result  in the present paper is the following.

\ms

\noindent
\begin{theorem}\label{thm1}If the obstacles $K, L\in \ll_0$ have almost the same travelling times,  then there exists a homeomorphism 
$\Phi : \dot{T}_b^*(\Omega_{K})\setminus \trapk  \longrightarrow  \dot{T}_b^*(\Omega_{L})\setminus\trapl$
which defines a symplectic map on an open dense subset of  $\dot{T}_b^*(\Omega_{K})\setminus \trapk$,
maps $S_b^*(\Omega_K)\setminus \trapk$ onto $S_b^*(\Omega_L)\setminus \trapl$,
and is such that $\fl_t\circ \Phi = \Phi\circ \fk_t$ for all $t\in \R$ and $\Phi = \mbox{id}$ on 
$\dot{T}_b^*(\oock)\setminus \trapk = \dot{T}_b^*(\oocl)\setminus\trapl$.
Conversely, if $K, L\in \ll_0$ are two obstacles for which there exists a homeomorphism 
$\Phi : S_b^*(\Omega_{K})\setminus \trapk  \longrightarrow S_b^*(\Omega_{L})\setminus\trapl$
such that $\fl_t\circ \Phi = \Phi\circ \fk_t$ for all $t\in \R$ and $\Phi = \mbox{id}$ on 
$S^*(\oock)\setminus \trapk$, then $K$ and $L$ have the same travelling times.
\end{theorem}

\ms

This result is similar to Theorem 1.1 in \cite{kn:St3}, and as a consequence of it we obtain 
analogues of all results in \cite{kn:St3} and \cite{kn:St2}. For example the existence of  the conjugacy $\Phi$ implies:

\ms

\noindent
\begin{corollary}\label{cor1} 
Let the obstacles $K,L\in \ll_0$ have almost the same travelling times. Then:
\begin{enumerate}
\item[{\rm (a)}]  If the sets of  trapped points of both $K$ and $L$ have Lebesgue measure zero, then $\Vo(K) = \Vo(L)$.

\item[{\rm (b})]   If  $K$ is star-shaped, then $L = K$.

\item[{\rm (c)}]  Let $\tkn$ be the set of those $x\in \dk$ such that $(x,\nu_K(x))\in \trapk$, where $\nu_K(x)$ is the {\it outward unit normal} to $\dk$ at $x$. 
There exists a homeomorphism \\ $\varphi: \:\dk\setminus \tkn \longrightarrow \dl\setminus \tln$ such that $\varphi(x) = y$ whenever 
$\Phi(x,\nu_K(x)) = (y,\nu_L(y))$.

\item[{\rm (d)}] Let $n \geq 3$, $\dim (\tkn) < n-2$ and $\dim (\tln) < n-2$. Then $K$ and 
$L$ have the same number of connected components.

\item[{\rm (e)}] Assume that $\dk$ and $\dl$ are both real analytic and $K$ is a finite union of 
strictly convex domains in $\R^n$. Then $K = L$.

\end{enumerate}
\end{corollary}

\ms

Very recently it was proved in \cite{kn:NS} that the conclusion in (e) remains true without assuming 
real analyticity of $\dk$ and $\dl$, namely if $K$ and $L$ are both disjoint unions of strictly convex
domains in $\R^n$ with $C^3$ boundaries and $K$ and $L$ have almost the same travelling times 
(or SLS), then $K = L$. This shows that, in principle, $K$ is determined by ${\cal T}_K$, and prompts the question of how to 
actually reconstruct $K$, assuming $K$ is a disjoint unions of $m$ strictly convex domains in $\R^n$ with smooth boundaries.
Except when $m=1$ (Example \ref{ex1}), this is not so simple to answer. 
The case $m=2$, $n = 2$ is dealt with in \S \ref{refsec}, \ref{twostr}. Different methods are needed for $m\geq 3$.

\section{Proof of Theorem \ref{thm1}}
\renewcommand{\theequation}{\arabic{section}.\arabic{equation}}

Let $\oo$ be the open ball with boundary $\partial \oo = S_0$. Set $\Omega = \R^n \setminus \oo$.   
As in \cite{kn:St3}, the main step in proving Theorem \ref{thm1} is to  show that the flows $\fk_t$ and $\fl_t$ coincide in $\Omega$.

\ms

\noindent
\begin{theorem}\label{thm2}
Let $K,L\in \ll_0$, If $K$ and $L$  have almost the same travelling times, then the flows $\fk_t$
and $\fl_t$ coincide on $S^*(\Omega)$. Namely, for every $\sigma\in S^*(\Omega)$ 
and every $t\in \R$ with $\fk_t(\sigma)\in S^*(\Omega)$  we have $\fk_t(\sigma) = \fl_t(\sigma).$
\end{theorem} 

\spp
The converse statement is also true: if $\fk_t(\sigma) = \fl_t(\sigma)$ holds for all  
$\sigma\in S^*(\Omega)$ and $t\in {\R}$ with $\fk_t(\sigma)\in S^*(\Omega)$, then 
$K$ and $L$ have (almost) the same travelling times. 

\ms

We prove Theorem \ref{thm1} using some ideas from \S 2 of \cite{kn:St3}. Some technical lemmas will be necessary. 
The first part of the following lemma  is an immediate consequence of Theorem 3.2 in \cite{kn:PS2}. The  second part follows from Sard's Theorem.

\ms

\noindent
\begin{lemma}\label{llem1} There exists a subset $\ss$ of $S^*_+(S_0)$ the complement of which 
is a countable union of compact subsets of Lebesgue measure zero in $S^*_+(S_0)$ such that
for every $(x,u) \in \ss$ the generalised geodesic $\gamma_K(x,u)$ in $S^*(\Omega_K)$ issued  from $x$ in 
direction $u$ has no tangencies to $\dk$ and moreover $\gamma_K(x,u)$ is regular.
\endofproof
\end{lemma}

\ms

The next lemma is similar to Proposition 4.1 in \cite{kn:PS2}. For completeness we sketch
its proof in the Appendix.

\ms

\noindent
\begin{lemma}\label{llem2} Let $x_0 \in S_0$ be a fixed point. There exists a subset $\ss(x_0)$ of $S^*(S_0)$ 
the complement of which  is a countable union of compact subsets of Lebesgue measure zero in $S_0$ such that
for every $y \in \ss(x_0)$ any two different $(x_0,y)$-geodesics in $\Omega_K$ that are simply reflecting
and regular, have distinct travelling times.
\end{lemma}

\ms

The final lemma we need is similar to Lemma 2.2 in \cite{kn:St3}.

\ms

\noindent
\begin{lemma}\label{llem3} Let $(x_0,y_0)\in S_0 \times S_0$, let $\gamma$ be a regular simply reflecting 
$(x_0,y_0)$-geodesic in $\Omega_K$ with successive reflection points $x_1, \ldots, x_k$ ($k\geq 1$), let $\omega_0 \in \sn$ be the incoming direction of $\gamma$ at $x_0$ and let
$\fk_{t_0} (x_0,\omega_0)= (y_0, \theta_0)$ for some $t_0 > 0$. Then there exist a neighbourhood $W$ of $(x_0,y_0)$ in $S_0 \times S_0$  and for each $i = 1, \ldots, k$ 
a neighbourhood $U_i$ of $x_i$ in $\dk$ such that  for any $(x,y)\in W$ there are unique $x_i (x,y) \in U_i$ ($i = 1, \ldots, k$) which are
the  successive reflection points of a simply reflecting  $(x,y)$-geodesic in $\Omega_K$. Moreover,
$x_i(x,y)$ depends smoothly on $(x,y) \in W$ for each $i = 1,\ldots,k$, and taking the neighbourhood 
$W$ sufficiently small, there exists an open neighbourhood $V$ of $(x_0,\omega_0)$ in  $S^*_+(S_0)$ such that the map 
$F(x,\omega) = (x,\pr_1(\pp_K(x,\omega)))$  defines a diffeomorphism $F : V \longrightarrow W$.
\end{lemma}

\ms

\noindent
{\bf Proof:} Let $\gamma$ be as above. Choosing a sufficiently small open neighbourhood $V$ of $\sigma_0 = (x_0, \omega_0)$ in $S^*_+(S_0)$,
the cross-sectional map  $\pp_K : V \longrightarrow S^*(S_0)$  is well-defined and smooth, and for every
$\sigma\in V$ the ray $\gamma_K(\sigma)$ is simply reflecting and has exactly $k$ reflection
points $x_1(\sigma), \ldots, x_k(\sigma)$, depending smoothly on $\sigma\in V$. Then $Y(x,\omega) = \pr_1 (\pp_K(x,\omega)) \in S_0$
defines a smooth map on $V$. Given $\omega$ close to $\omega_0$, let $Y_{\omega} = Y(.,\omega)$.
Then $\det dY_{\omega}(x)$ depends smoothly on $(x,\omega)\in V$, and, since $\gamma$ is regular, 
we have $\det dY_{\omega_0} (x_0) \neq 0$. Assuming $V$ small enough, we have  $\det dY_{\omega} (x) \neq 0$ for all $(x,\omega) \in V$.
Setting $F(x,\omega) = \left( x , \pr_1 (\pp_K(x,\omega)) \right) = (x, Y(x,\omega))$, we get
a smooth map $F : V \longrightarrow  S_0\times S_0$ whose differential at $(x,\omega)\in V$ has the form
$$ dF (x, \omega) = \pmatrix{I              & 0 \cr 
                                               *             & dY_{\omega}(x) \cr}\; ,$$
where $I$ is the identity $(n-1)\times (n-1)$ matrix. Thus, $\det dF (x,\omega) \neq 0$. By the inverse 
mapping theorem, $F$ is locally invertible. Shrinking $V$ if necessary, 
$F$ is a diffeomorphism between $V$ and an open neighbourhood $W$ of $(x_0,y_0)$ in $S_0 \times S_0$. 
Setting $G = F^{-1}$, we have $G(x,y) = (x , \omega(x,y))$ for $(x,y)\in W$, where 
$\omega(x,y)$ is so that $Y(x,\omega(x,y)) = y$. Then  $x_j(x,y) = x_j (x , \omega(x,y) )$ 
depends smoothly on $(x,y)\in W$ for all $j = 1, \ldots,k$, and clearly $x_1 (x,y), \ldots, x_k (x,y)$ 
are the successive reflection points of a simply reflecting $(x,y)$-geodesic in $\Omega_K$. Finally, if $y_1,\ldots,y_k$ are the successive reflection 
points of a $(x,y)$-geodesic in $\Omega_K$ for some $(x,y) \in W$ and  $y_j\in \dk$ is sufficiently close to $x_j(x_0,y_0)$ for all $j$, then 
we must have $y_j = x_j (x,y)$ for all $j = 1,\ldots,k.$ 
\endofproof

\ms

Given two obstacles $K$ and $L$  in ${\R}^n$, fix for a moment a pair of points  $(x_0,y_0)\in S_0\times S_0$. Let $\delta$ be a 
non-degenerate simply reflecting $(x_0,y_0)$-geodesic in $\Omega_K$ with reflection points $x_1,\ldots,x_k$
($k \geq 1$) and let $\delta'$ be a non-degenerate simply reflecting $(x_0,y_0)$-geodesic in $\Omega_L$ with
reflection points $y_1,\ldots,y_m$ ($m \geq 1$). By Lemma \ref{llem3}, there exists a
neighbourhood $W$ of $(x_0,y_0)$ in $S_0 \times S_0$ such  that for each $(x,y)\in W$ there are a unique reflecting
$(x,y)$-geodesic $\delta (x,y)$ in $\Omega_K$ with  reflection points $x_1(x,y),\ldots,x_k(x,y)$ close to
$x_1,\ldots, x_k$, and a unique   reflecting $(x,y)$-geodesic $\delta'(x,y)$ in $\Omega_L$ with
reflection points $y_1(x,y),\ldots,y_m(x,y)$ close  to $y_1,\ldots,y_m$.

An important step in the proof of Theorem 2 is the following.

\ms

\begin{lemma}\label{llem5}
Under the above assumptions, suppose in addition that $t_{\delta(x,y)} = t_{\delta'(x,y)}$ for all 
$(x,y)\in W$. Then for each $(x,y)\in W$ there exist $\omega(x,y), \theta(x,y) \in \sn$ such that
\begin{equation}
\omega(x,y) = \frac{x_1(x,y) - x}{\|x_1(x,y) - x\|} = \frac{y_1(x,y) - x}{\|y_1(x,y) - x\|} 
\end{equation}
and
\begin{equation}
\theta(x,y) = \frac{y- x_k(x,y)}{\|y- x_k(x,y) \|} = \frac{y- y_k(x,y)}{\|y - y_k(x,y)\|} .
\end{equation}
In other words the $(x,y)$-geodesics $\delta(x,y)$ and $\delta'(x,y)$ income through $x$ with the same
direction $\omega(x,y)$ and outgo through $y$ with the same direction $\theta(x,y)$.\end{lemma}

\ms

{\bf Proof:} Let $(x(u),y(v))$ be a smooth parametrisation of $W$ near $(x_0,y_0)$ with
$u \in U$, $v \in V$ for some open subsets $U$ and $V$ of $\R^{n-1}$. Set 
$x_i(u,v) = x_i(x(u),y(v))$ for $i = 1, \ldots,k$ and $y_j(u,v) = y_j(x(u),y(v))$ for $j =1, \ldots, m$,  $x_0(u,v) = y_0(u,v) = x(v)$,
$x_{k+1}(u,v) = y_{m+1}(u,v) = y(v)$. We will also need the unit vectors
$$e_i(u,v) = \frac{x_{i+1}(u,v) - x_i(u,v)}{\|x_{i+1}(u,v) - x_i(u,v)\|} \quad, \quad
\te_j(u,v) = \frac{y_{j+1}(u,v) - y_j(u,v)}{\|y_{j+1}(u,v) - y_j(u,v)\|} $$
for all $i = 0,1, \ldots,k$ and $j = 0,1, \ldots, m$. 
 
Consider the smooth functions
$$f(u,v) = \sum_{i=0}^k \| x_i(u,v) - x_{i+1}(u,v)\|  \quad , \quad
g(u,v) =  \sum_{j=0}^m \| y_j(u,v) - y_{j+1}(u,v)\| .$$
Since $f(u,v) = g(u,v)$ for all $(u,v) \in U\times V$, the derivatives of these two functions coincide. We have
\begin{eqnarray*}
\frac{\partial f}{\partial u_j} (u,v) 
& = &  \sum_{i=0}^{k} \left\langle \frac{x_{i}-x_{i+1}}{\|x_{i}-x_{i+1}\|},\frac{\partial x_{i}}{\partial u_j}  - \frac{\partial x_{i+1}}{\partial u_j}\right\rangle
= - \sum_{i=0}^{k} \left\langle e_i , \frac{\partial x_{i}}{\partial u_j}  - \frac{\partial x_{i+1}}{\partial u_j}\right\rangle\\
& = & - \left\langle  e_0 , \frac{\partial x_0}{\partial u_j} \right\rangle  - \left\langle e_1 - e_0 , \frac{\partial x_1}{\partial u_j} \right\rangle 
- \ldots  \cr
&   & - \left\langle e_{k} - e_{k-1},\frac{\partial x_{k}}{\partial u_j} \right\rangle +
\left\langle e_{k} ,\frac{\partial x_{k+1}}{\partial u_j} \right\rangle
= - \left\langle e_0(u,v) , \frac{\partial x}{\partial u_j}(u) \right\rangle ,
\end{eqnarray*}
where we took into account that $e_j- e_{j-1}$ is a normal vector to $\dk$ at $x_j(u,v)$, while $\frac{\partial x_{j}}{\partial u_j}(u,v)$
is tangent to $\dk$ at the same point, and also that $x_{k+1}(u,v) = y(v)$ does not depend on $u$.

Similarly, using the fact that $y_0(u,v) = x_0(u,v)$, one gets 
$$\displaystyle \frac{\partial g}{\partial u_j} (u,v)=  - \left\langle \te_0(u,v) , \frac{\partial x}{\partial u_j}(u) \right\rangle .$$
Hence
$$\displaystyle  \left\langle \te_0(u,v) - e_0(u,v) , \frac{\partial x}{\partial u_j} (u) \right\rangle =  0$$
for all $j = 1,\ldots,n-1$. Since  the vectors $\frac{\partial x}{\partial u_j} (u)$ ($j = 1, \ldots, n-1$) form a basis in the 
tangent space to $S_0$ at $x(u)$, this shows that $\te_0(u,v) - e_0(u,v)$ is parallel to the inward unit normal vector $\nu(x_0(u,v))$ to $S_0$
at $x_0(u,v)$. Since $e_0(u,v)$ and $\te_0(u,v)$ are unit vectors, it now follows that\footnote{Indeed, fix for a moment
$u$ and $v$. Considering an appropriate coordinate system in $\R^n$, we may assume that $\nu(x_0(u,v)) = -e_n = (0, \ldots, 0,-1)$. Let 
$e_0(u,v) = (x_1, \ldots, x_{n-1}, x_n)$ and $\te_0(u,v) = (y_1, \ldots, y_{n-1}, y_n)$, where
obviously we must have $x_n < 0$ and $y_n < 0$. Then $e_0(u,v) - \te_0(u,v) = \lambda \, e_n $ for some $\lambda\in \R$,
so $x_i = y_i$ for all $i = 1, \ldots, n-1$. Now $\|e_0(u,v)\| = \|\te_0(u,v)\| = 1$ and $x_n y_n > 0$ imply $e_0(u,v) = \te_0(u,v)$.}
$e_0(u,v) = \te(u,v)$ for all $(u,v) \in U\times V$.

In a similar way, differentiating with respect $v_j$, we get $e_k(u,v) = \te_k(u,v)$ for all $(u,v) \in U\times V$.
\endofproof

\bs

\spp
The proof of the lemma shows that from the derivatives of the travelling time function $f(u,v)$, under the non-degeneracy
assumption of the lemma, we can recover the { incoming and outgoing directions} $\omega(u,v)$ and $\theta(u,v)$
of the $(x(u),y(v))$-geodesics in $\Omega_K$. Some other constructive ideas are discussed in \S \ref{refsec} below.

$\:$\\
{\bf Proof of Theorem \ref{thm2}:} We follow the main steps in the proof of Theorem 2.1 in \cite{kn:St3} with necessary modifications.

Assume that two obstacles  $K, L\in \ll_0$  have almost the same travelling times.
Thus, there exists a subset $\rr$ of full Lebesgue measure in $S_0 \times S_0$ such that
\begin{equation}
\tt_K (x,y)= \tt_L(x,y) \quad , \quad   (x,y) \in \rr .
\end{equation}
As in Example 1, it follows that for $(x,y)\in \rr$, if a $(x,y)$-geodesic in $\Omega_K$ has  a common point with $\dk$,
then any $(x,y)$-geodesic in $\Omega_L$ has a common point with $\dl$ (and vice versa).
Using Lemma \ref{llem2} above, we may assume $\rr$ is chosen so that:
(i) for $(x,y)\in \rr$ all $(x,y)$-geodesics in $\Omega_K$ and $\Omega_L$ having at least one common point with $\dk$ and $\dl$,
respectively, are regular  and simply reflecting; 
(ii) for all $(x,y)\in \rr$ we have $t_\gamma \neq t_\delta$ whenever $\gamma$ and $\delta$ are regular simply reflecting 
$(x,y)$-geodesics in $\Omega_K \cap \oo$ (or $\Omega_L \cap \oo$) each of them having at least one reflection point.

To prove the theorem it is enough to show that for every $\sigma\in S^*_+(S_0)$ 
and every $t > 0$ with $\fk_t(\sigma)\in S^*(S_0)$  we have $\fk_t(\sigma) = \fl_t(\sigma).$

Let $\ts = (\tilde{x}_0,\tilde{\omega}_0)\in S^*_+(S_0)$ and $t_0 > 0$ be such that  $\fk_{t_0}(\ts) \in S^*(S_0)$. We are going to show that  $\fk_{t_0}(\ts) = \fl_{t_0}(\ts)$. 
It follows from  Proposition 2.3 of \cite{kn:St2} that $S^*_+(S_0) \setminus \SL$ has Lebesgue measure zero in  
$S^*_+(S_0)$.  Since the flows $\fk_{t_0}$ and $\fl_{t_0}$ are both continuous, it is enough to consider the case $\ts \in \SK$ and $\ts \in \SL$. 
Similarly, using Lemma \ref{llem2}, we may assume that $\gamma_K(\ts)$ and $\gamma_L(\ts)$ are simply reflecting regular rays each of them having at least one reflection point.

Let $\fk_{t_0} (\ts) = (\tilde{y}_0, \tilde{\theta}_0)$; then $\tilde{y}_0 \in S_0$ and  $\gamma_K(\ts)$ is a  $(\tilde{x}_0,\ty_0)$-geodesic.  
Using the diffeomorphism $F$ from Lemma \ref{llem5} and the fact that $\rr$ is dense in $S_0\times S_0$ 
we can find  $\sigma_0 = (x_0,\omega_0) \in S^*_+(S_0)$  arbitrarily close to $\ts$ such that for 
$\pp_K(x_0,\omega_0) = (y_0,\theta_0)$ we have $(x_0,y_0) \in \rr$.

Let $x_1, x_2, \ldots, x_k$ be the successive reflection points of $\delta = \gamma_K(x_0, \omega_0)$, where $k \geq 1$.
By (2.3), there exists a (simply reflecting, regular)  $(x_0,y_0)$-geodesic $\delta'$ in $\Omega_L$  with  
\begin{equation}
t_{\delta'} = t_{\delta} .
\end{equation}
If  $y_1,\ldots,y_m$ are the successive reflection points of $\delta'$, we must have $m \geq 1$ (see e.g. Example 1).
Using  Lemma \ref{llem3}, there exist a neighbourhood $W$  of $(x_0,y_0)$ in $\ssn$ and a 
neighbourhood $U_i$ of $x_i$ in $\dk$ for each $i = 1,\ldots,k$ such that for every $(x,y)\in W$ 
there is a unique reflecting $(x,y)$-geodesic $\delta(x,y)$ in $\Omega_K$ with reflection points 
$x_i (x,y)\in U_i$ ($i = 1, \ldots,k$) smoothly depending on $(x,y)$. By the same argument, there exists a
neighbourhood $U'_j$ of $y_j$ in $\dl$ for each $j = 1, \ldots, m$ such that for every $(x,y)\in W$ 
there is a unique reflecting $(x,y)$-geodesic $\delta'(x,y)$ in $\Omega_L$ with reflection 
points $y_i(x,y)\in U'_i$ ($i=1, \ldots,m$) smoothly depending on $(x,y)$. Clearly
$\delta(x_0,y_0) = \delta$ and $\delta'(x_0,y_0) = \delta'$.

Next, using (2.3) again, for $(x,y)\in \rr\cap W$ there exists a unique reflecting $(x,y)$-geodesic $\delta''(x,y)$ in $\Omega_L$ with
\begin{equation}
t_{\delta''(x,y)} = t_{\delta(x,y)}.
\end{equation}
We claim that if $W$ is chosen small enough, then $\delta''(x,y) = \delta'(x,y)$ for all $(x,y)\in \rr\cap W$. 
Assume this is not true. Then  there exists a sequence  $\{ (x_p,y_p)\}_{p=1}^\infty\subset \rr\cap W$ with $(x_p,y_p) \to (x_0,y_0)$ 
as $p\to \infty$ and $\delta''(x_p,y_p) \neq \delta'(x_p,y_p)$ for all $p$. 
Denote by $\omega_p$ the (incoming) direction of $\delta''(x_p,y_p)$ at $x_p$;  then $\delta''(x_p,y_p) = \gamma_L(x_p,\omega_p)$. 
Taking a subsequence, we may assume $\omega_{p} \rightarrow \omega\in \sn$ as 
$p\to \infty$. Then $\delta'' = \gamma_L(x_0, \omega)$ is a $(x_0,y_0)$-geodesic in $\Omega_L$ and clearly
$t_{\delta''} = \lim_p t_{\delta''(x_p,y_p)} = t_{\delta''(x_0,y_0)}$.  By (2.5), $t_{\delta''} = t_{\delta(x_0,y_0)} = t_\delta$, and by (2.4),
$t_{\delta''} = t_{\delta'}$. Now $(x_0,y_0)\in \rr$ implies $\delta'' = \delta'$. Thus, $x$ lies on
$\delta' = \delta'(x_0,y_0)$, so for large $p$, the ray $\delta''(x_p,y_p)$ has $m$
reflection points the $j$th of which is in $U'_j$. Now the choice of $W$ and the uniqueness of the $(x,y)$-geodesics $\delta'(x,y)$ for $(x,y)\in W$ imply
$\delta'' (x_p,y_p) = \delta'(x_p,y_p)$. This is impossible by the choice of the sequence $\{ (x_p,y_p) \}_p$ which proves that $\delta''(x,y) = \delta'(x,y)$
for all $(x,y)\in \rr\cap W$. Therefore
\begin{equation}
t_{\delta'(x,y)} = t_{\delta(x,y)}
\end{equation}
for $\ot \in \rr\cap W$. Since $\rr\cap W$ is dense in $W$, by continuity  it follows that (2.6) holds for all $(x,y) \in W$. 
Now Lemma \ref{llem5} shows that $\delta'(x,y)$ and $\delta(x,y)$ have the same incoming direction at $x$
and the same outgoing direction at $y$. In particular, $\delta' = \gamma_L(\sigma_0)$ and it has outgoing direction $\theta_0$ at $y_0$. Thus, 
$\fl_{t_{\delta'}}(\sigma_0) = (y_0,\theta_0) = \fk_{t_{\delta}}(\sigma_0)$.
Letting $\sigma_0$ tend to $\ts$ (and then $t_{\delta} = t_{\delta'}$ tends to $t_0$), we get $\fl_{t_0}(\ts) = \fk_{t_0}(\ts)$.
\endofproof

\bs

\noindent
{\bf Proof of Theorem \ref{thm1}:} This is now done exactly as the proof of
Theorem 1.1 in \cite{kn:St3}.
\endofproof

%
\section{Reflexive Geodesics}\label{refsec}
For $m\geq 1$ let $K$ be the union of $m$ disjoint strictly convex bodies $K_1,K_2,\ldots ,K_m$ whose boundaries are 
$C^\infty$ manifolds of dimension $n-1$.
Then, as indicated following Corollary \ref{cor1} and proved in \cite{kn:NS}, $K$ is determined uniquely by ${\cal T}_K$. 
The proof in \cite{kn:NS} is nonconstructive, and it is not easy to 
give a constructive algorithm except in simple cases. The case where $m=1$ is easily dealt with in Example \ref{ex1}. 
The present section and the next give a construction sufficient for $m=2$.

\spp
 For $(x,y)$ in some subset $U$ of $S_0\times S_0$ whose complement is a 
countable union of compact subsets of Lebesgue measure zero, all $(x,y)$-geodesics are regular and simply reflecting. 
Given an $(x_0,y_0)$-geodesic $\gamma _0$ where $(x_0,y_0)\in U$, there is a connected open subset $U_{\gamma _0 }$  of 
$S_0\times S_0$ containing $(x_0,y_0)$ and a unique $C^\infty$ assignment $(x,y)\in U_{\gamma _0}\mapsto \gamma _{(x,y)}$ with 
$\gamma _{(x_0,y_0)}=\gamma _0 $ such that 
$\gamma _{(x,y)}$ is a regular simply reflecting $(x,y)$-geodesic. Take $U_{\gamma _0}$ to be the largest such subset and  
define  
$\displaystyle{{\cal T}_{\gamma _0}(x,y):=t_{\gamma _{(x,y)}}}$. Then ${\cal T}_{\gamma _0}:U_{0}\rightarrow \R$ is $C^\infty$ and  
${\cal T}_{\gamma _0}(x,y)\in {\cal T}_K(x,y)$. The following result is proved in the same way as Lemma \ref{llem5}.
\begin{lemma}\label{lem1} For $a$ and $b$ tangent to $S_0$ at $x$ and $y$ respectively where $(x,y)\in U_{0}$, we have 
$$d{{\cal T}_{\gamma _0}}_{(x,y)}(a,b)=\langle v,b\rangle -\langle u,a\rangle $$ 
where $u$ and $v$ are the unit vectors in the directions of $\dot \gamma _{x,y}$ at $x$ and 
$y$ respectively. \endofproof
\end{lemma}

\spp
In particular $u$ and $v$ are found by differentiating ${\cal T}_{\gamma _0}$.

\begin{definition}
An $(x,y)$-geodesic $\gamma $ is said to be {\em reflexive} 
when it intersects some point of $\partial K$ orthogonally. Then $x=y$ and, for some $p\geq 1$,  
the points of intersection of $\gamma$ can be written $z_1,z_2,\ldots ,z_{p-1},z_p,z_{p-1},\ldots ,z_2,z_1$. The integer $p$ is the {\em order} of the reflexive geodesic $\gamma $. \endofproof
\end{definition} 
Reflexive geodesics $\gamma $ correspond precisely to 
generalised geodesics leaving $K$ orthogonally at $z_p$, then intersecting $K$ at $z_{p-1},z_{p-2},\ldots ,z_2,z_1$ before ending on 
$S_0$. We call $z_p$ the {\em middle reflection point} of the  reflexive geodesic. 
When $z\in \partial K$ is $z_p$ for a reflexive geodesic $\gamma $, the {\em order} of $z$ is defined to be the order of $\gamma $. Any $z\in \partial K$ that is not the end of a reflexive geodesic is said to have {\em infinite order}.  
The set of points of finite order is the complement in $\partial K$ 
of a countable union of compact subsets of Lebesgue measure zero. 

\spp 
Examples  
can be given of non-reflexive $(x,x)$-geodesics. When $(x,x)\in U_{0}$, a necessary and sufficient condition for 
$\gamma _{(x,x)}$ to be reflexive is that $v=-u$. This condition can be tested by differentiating 
${\cal T}_{\gamma _0}$ as in Lemma \ref{lem1}.

\spp
Given a reflexive $(x_0,x_0)$-geodesic $\gamma _0$ where $(x_0,x_0)\in U$, let $V_{\gamma _0}$ be the largest connected open subset of 
$S_0$ with the property that $\gamma _{(x,x)}$ is reflexive for all 
$x\in V_{\gamma _0}$.  Then for $x\in V_{\gamma _0}$ the middle reflection point of $\gamma _{(x,x)}$ has constant order, namely the order $p$ of $\gamma _0$. 
For $x\in V_{\gamma _0}$ define ${\cal S}_{\gamma _0}(x)={\cal T}_{\gamma _0}(x,x)$. Then   ${\cal S}_{\gamma _0}:V_{\gamma _0}\rightarrow \R$ is $C^\infty$ and from   
Lemma \ref{lem1} (or Lemma \ref{llem5}) we obtain the following result.  
\begin{lemma}\label{lem1.5} For $x\in V_{\gamma _0}$,  the unit vector in the direction of $x- z_1$ is determined by the 
derivative of ${\cal S}_{\gamma _0}$ at $x\in V_{\gamma _0}$. \endofproof
\end{lemma}
In particular when $\gamma _0$ has order $1$,   
${\cal S}_{\gamma _0}(x)=2\Vert x-z_1\Vert $. Consequently 
\begin{lemma}\label{lem2} The set $Z$ of points of order $1$ is determined by the real-valued functions  ${\cal S}_{\gamma _0}$ defined by first order 
reflexive $(x_0,x_0)$-geodesics $\gamma _0$ where $(x_0,x_0)\in U$ . \endofproof
\end{lemma}

\spp
It remains to construct the ${\cal S}_{\gamma _0}$ from the set-valued ${\cal T}_K$. Here 
we do so only for the following special case, where $\gamma _0$ has order $1$ and $K$ has $2$ connected components.

\begin{example}\label{m=2ex}
Set $m=2$. Then the infimum $\tau _1$ of $\cup _{x\in S_0}{\cal T}_K(x,x)$ lies in     
${\cal T}_K(x_0,x_0)$ for some (at most two) $x_0\in S_0$. The inward ray orthogonal to $S_0$ at 
$x_0$ meets $\partial K$ orthogonally at $z_1=x_0+\tau _1\nu _{x_0}/2$, and the interval $x_0z_1$ does not meet $K$ at any other point. So the track-sum of $x_0z_1$ 
and $z_1x_0$ is an order $1$ reflexive geodesic $\gamma _0$, which can be found 
from ${\cal T}_K$. Set $y_0=x_0$ and define $U_{\gamma _0}$ as before. 
Then ${\cal T}_{\gamma _0}$ determines ${\cal S}_{\gamma _0}$, and for all $(x,y)\in U_{\gamma _0}$ there exists an $(x,y)$-geodesic meeting $\partial K$ 
only on $ \partial K_1$, where the $K_i$ are labelled so that $z_1\in \partial K_1$.  

\spp
The space $V$ of vacuous hyperplanes is determined from ${\cal T}_K$ as in \S \ref{sec1}, and has two path-components because $m=2$. 
Choose a vacuous hyperplane $H$ in the path component of $V$ that does not contain the trivially vacuous hyperplanes. 
Then $K_2$ is contained in the interior of the closed half-space $\tilde H$ bounded by $H$ for which  
$z_1\notin \tilde H$.

\spp
Because $K_2$ is strictly convex, the infimum $\tau _2$ of $\cup _{x\in S_0\cap \tilde H}{\cal T}_K(x,x)$ 
lies in ${\cal T}_K(x_2,x_2)$ 
for some $x_2\in S_0\cap \tilde H$. The inward ray orthogonal to $S_0$ at $x_2$ meets $\partial K_2$ 
orthogonally at $z_2=x_2+\tau _2\nu _{x_2}/2\in \partial K_2$, and $x_2z_2$ does not meet 
$\partial K$ at any other point. So 
 $x_2z_2$ and $z_2x_2$ define an order $1$ reflexive geodesic $\gamma _2$. Define $U_{\gamma _2}$ and 
 ${\cal T}_{\gamma _2}:U_{\gamma _2}\rightarrow \R$ by analogy with $U_{\gamma _0}$ and ${\cal T}_{\gamma _0}$. 
 Then ${\cal T}_{\gamma _2}$ determines ${\cal S}_{\gamma _2}$, and for all $(x,y)\in U_{\gamma _2}$ there exists an $(x,y)$-geodesic meeting $\partial K$ 
only on $ \partial K_{\gamma _2}$. \endofproof
\end{example}

\spp
In \S \ref{sec1} a subset $Y$ of $\partial K$ is constructed  using vacuous geodesics found from ${\cal T}_K$. Sometimes $Y\cup Z$ is 
dense in $\partial K$, but usually not when the $K_i$ are close together. When $m=2$ we can recover  
$\partial K$ as follows.
\section{Two Strictly Convex Components}\label{twostr} 
Take $m=2$ and set $\rho _K:=\min \{ t:(x,x,t)\in {\cal T}_K\}$. By strict convexity  
$X_K:=\{ x\in S_0:(x,x,\rho _K)\in {\cal T}_K\}$ has at most two points; we consider the generic case where $X_K$ is a singleton 
$\{ x_K\}$. Then $z_K:=x_K+
(\rho _K/2)\nu _0(x_K)\in \partial K$. 
We label the components of $K$ so that $z_K\in \partial K_1$.  

\begin{definition}
Define ${\cal Z}_0=\emptyset $ and, for $k\geq 1$, let ${\cal Z}_k$ be the set of all $z\in \partial K$ 
such that the open outward normal ray from $z$ first meets $K$ only in ${\cal Z}_{k-1}$. $\endofproof$
\end{definition}
In particular $x_K\in {\cal Z}_1$. The ${\cal Z}_k$ are open and mutually disjoint subsets of $\partial K$, 
alternatively characterized as follows.

\spp
Denote the unit outward normal field on $\partial K$ by $\nu $, and let $\gamma _z$ 
be the generalized geodesic beginning at $z\in \partial K$ with initial direction $\nu (z)$. Let $\pi (z)$ be the first intersection after $z$ of $\gamma _z$ with $S_0\cup \partial K$.  Then ${\cal Z}_1=\pi ^{-1}S_0$ and 
${\cal Z}_{k+1}=\pi ^{-1}{\cal Z}_k$ for $k\geq 1$. 

\begin{figure}[h] 
   \centering
   \includegraphics[width=3in]{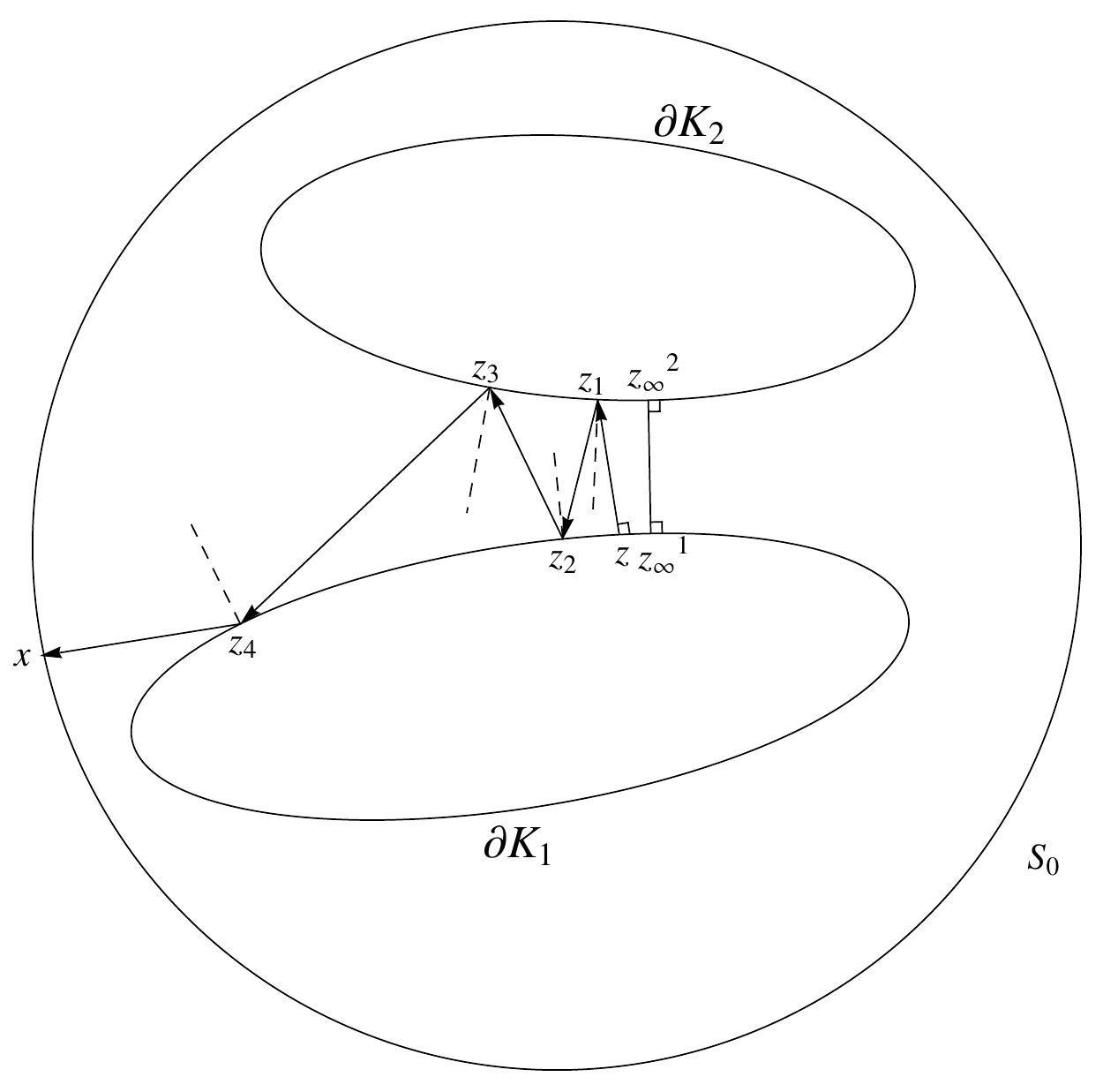} 
   \caption{$z_1=\pi (z)$ where $z\in {\cal Z}_k^{L,1}$ and $p=4$}
   \label{fig1}
\end{figure}

\spp
We have ${\cal Z}_1={\cal Z}_1^1\cup {\cal Z}_1^2$ where ${\cal Z}_1^i$ is an open 
arc in $\partial K_i$ for $i=1,2$. 
Let $z_{\infty}^1z_{\infty}^2$ 
be the shortest line segment joining $K_1$ and $K_2$, where $z_{\infty}^i\in \partial K_i$.  
After rotating $z_{\infty}^1z_{\infty}^2$ to   
vertical, the endpoints of the ${\cal Z}_1^i$ are labelled as left or right, namely $z_1^{L,i}$ and $z_1^{R,i}$. 
\begin{figure}[t] 
   \centering
   \includegraphics[width=3in]{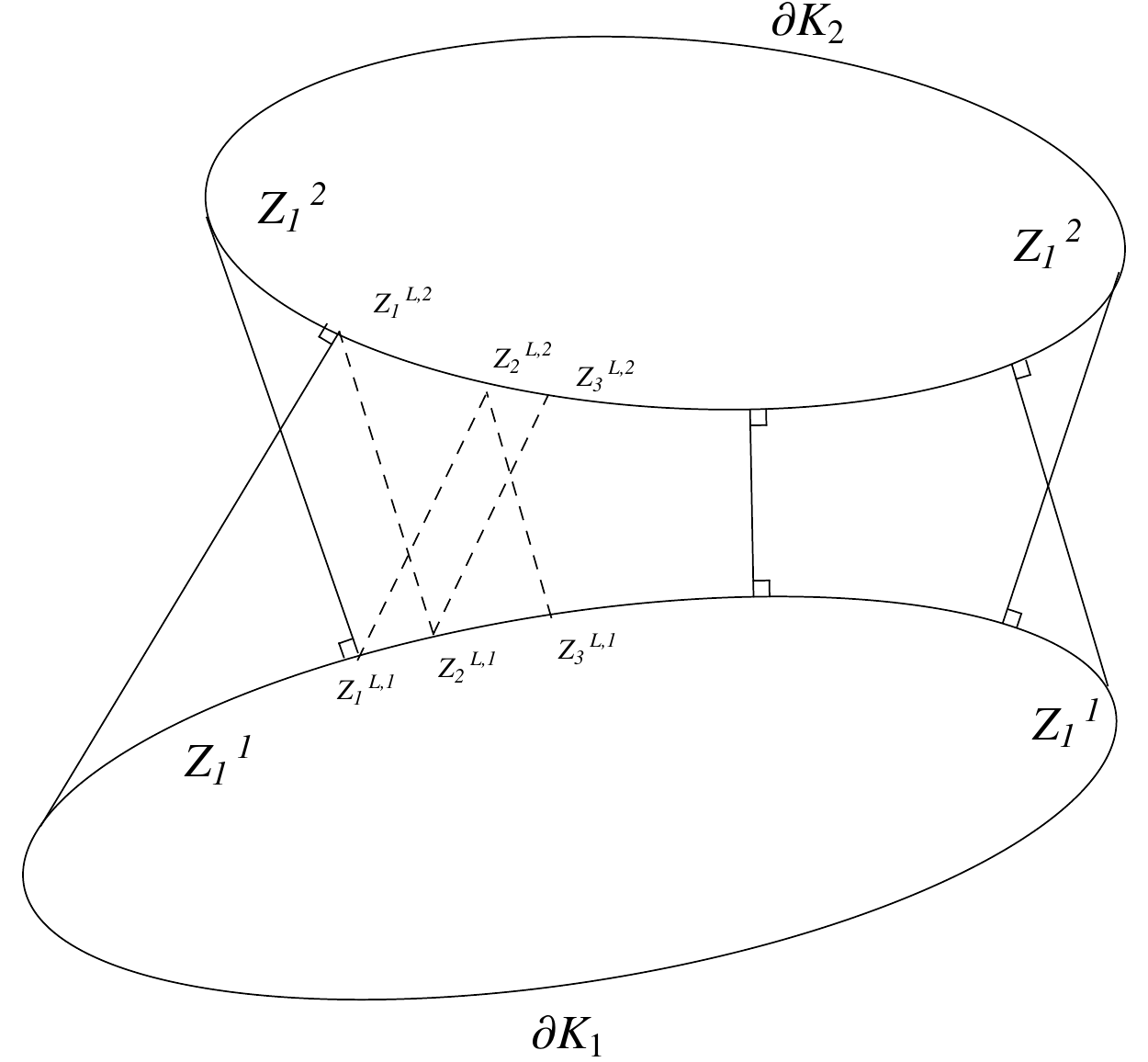} 
   \caption{$z_{k}^{L,1}z_{k-1}^{L,2}$ orthogonal to $\partial K_1$ ~and~ $z_k^{L,2}z_{k-1}^{L,1}$ orthogonal to $\partial K_2$ }
   \label{fig1.5}
\end{figure}

\spp
For $k\geq 2$, ${\cal Z}_k$ has two path components on each side of $z_{\infty}^1z_{\infty}^2$, labelled 
(left and right) thus: 
$${\cal Z}_k={\cal Z}_k^L\cup {\cal Z}_k^R\hbox{~~where~~}{\cal Z}_k^L={\cal Z}_k^{L,1}\cup {\cal Z}_k^{L,2}\hbox{~~and~~}{\cal Z}_k^R
={\cal Z}_k^{R,1}\cup {\cal Z}_k^{R,2}$$
with ${\cal Z}_k^{L,i},{\cal Z}_k^{R,i}\subset \partial K_i$ mutually disjoint open arcs. Then  
${\cal Z}_1^i$ is separated from ${\cal Z}_2^{L,i}$ only by $z_1^{L,i}$ and,  
for $k\geq 2$, 
${\cal Z}_{k}^{L,i}$ is separated from ${\cal Z}_{k+1}^{L,i}$ by a single point $z_k^{L,i}$.  So  
${\cal Z}_k^{L,i}$ is the open arc in $\partial K_i$ from $z_{k-1}^{L,i}$ to $z_{k}^{L,i}$ and 
$\lim_{k\rightarrow \infty}z_k^{L,i}=z_\infty ^i$, as illustrated in Figure \ref{fig1.5}.

\spp
For uniformity of notation, write ${\cal Z}_1^{L,i}:={\cal Z}_1^{R,i}:={\cal Z}_1^{i}$ where $i=1,2$. 
\begin{lemma}\label{backlem} For  $z\in {\cal Z}_k^{L,i}$ where $k\geq 2$ and $i=1,2$, all reflection points of $\gamma _z$ lie in  
$\cup _{j<k}({\cal Z}_j^{L,1}\cup {\cal Z}_j^{L,2})$.   
\end{lemma}

\spp
{\bf Proof:} Suppose $z\in {\cal Z}_k^{L,1}$, and let $z_1,z_2,\ldots ,z_p\in \partial K$ be the reflection points, taken in order from $z$, of $\gamma _z$ as in Figure \ref{fig1}. Then $z_q$ lies in $\partial K_1$ or $\partial K_2$ according as $q$ is even or odd, where $1\leq q\leq p$ and $1\leq p\leq k-1$. From the definition of ${\cal Z}_k^{L,1}$ we have $z_1\in {\cal Z}_{k-1}^{L,2}$. If $k=2$ this proves the lemma. 

\spp
If $k>2$ suppose inductively that, for some $1\leq q<p$ and all $1\leq r\leq q$ we have 
$z_r\in {\cal Z}_{j_r}^{L,1}\cup {\cal Z}_{j_r}^{L,2}$ where $k>j_1>j_2>\ldots >j_q> 1$. \newline 
If $q$ is even then $z_{q-1}\in {\cal Z}_{j_{q-1}}^{L,2}$, $z_q\in {\cal Z}_{j_q}^{L,1}$. 
From the definition of 
${\cal Z}_{j}^{L,i}$, the unit outward normal $\nu (z_q)$ from $K_1$ at $z_q$ points towards 
${\cal Z}_{j_q-1}^{L,2}$, namely $\pi (z_q)\in {\cal Z}_{j_q-1}^{L,2}$. The line segment $z_{q-1}z_q$ approaches $z_q$ from the right, and so its reflection meets $\partial K_1$ at least as far to the left as $\pi (z_q)$, 
namely $z_{q+1}\in  {\cal Z}_{j_{q+1}}^{L,1}$
where $j_{q+1}<j_q$. If $q$ is odd a similar argument holds with superscripts $1$ and $2$ interchanged.  In either case $z_{q+1}\in  {\cal Z}_{j_{q+1}}^{L,1}\cup {\cal Z}_{j_{q+1}}^{L,2}$, which completes the induction. So for all $1\leq q\leq p$, we have $z_q\in {\cal Z}_{j_q}^{L,1}\cup {\cal Z}_{j_q}^{L,2}$ with $1\leq j_q<k$, where the sequence $\{ j_q:1\leq q\leq p\}$ is strictly decreasing. This proves the lemma. 
$\endofproof$

\begin{figure}[h] 
   \centering
   \includegraphics[width=6.0in]{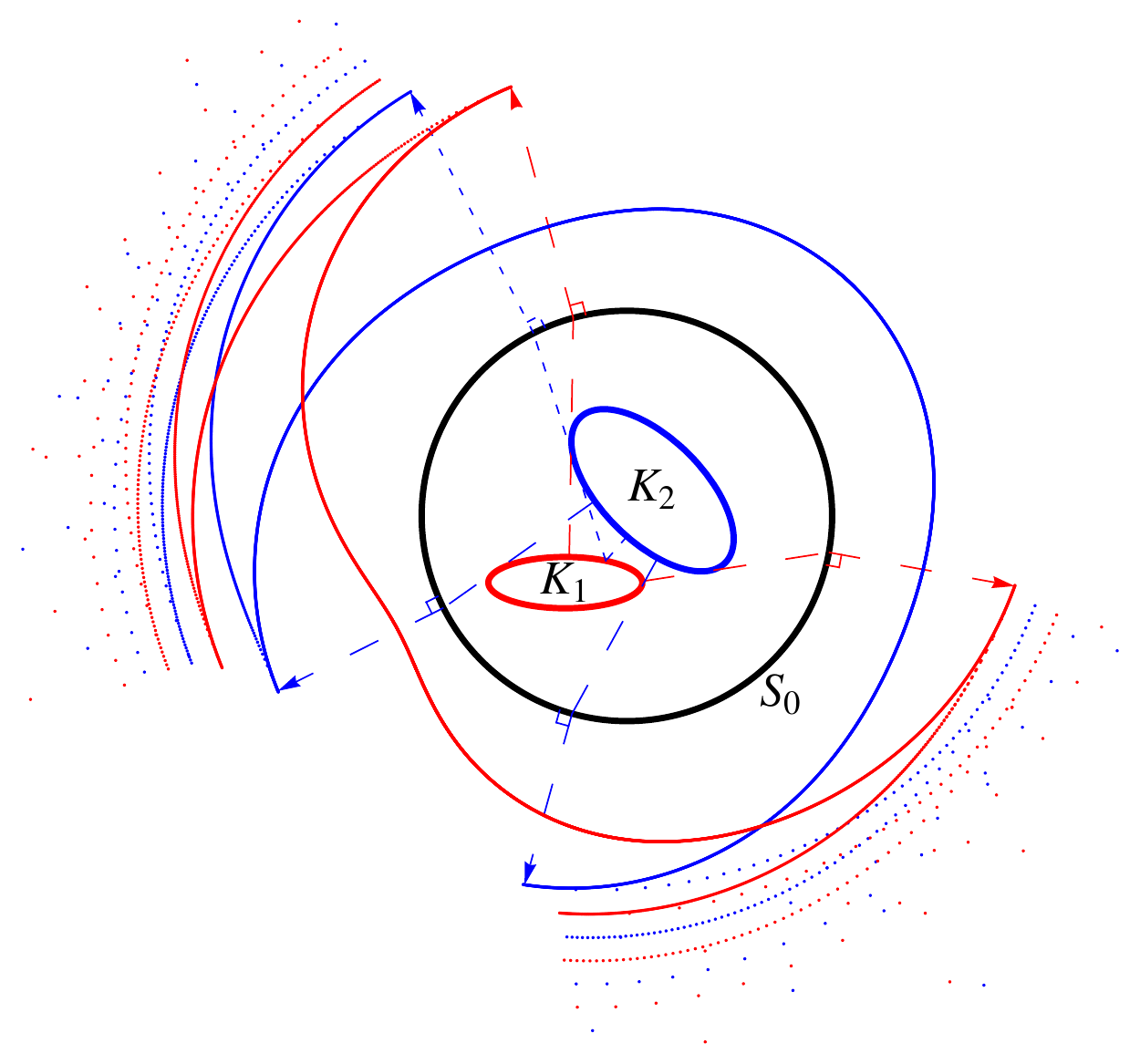} 
   \caption{${\cal E}_K$ for Example \ref{ex11}.}
   \label{echofig}
\end{figure}

\spp
The set $Z$ found from Example \ref{m=2ex} is ${\cal Z}$ whose path-components are ${\cal Z}_1^1$ and ${\cal Z}_1^2$. To find the rest of $K$ we need  ${\cal Z}_{k}^{L,i}$ and ${\cal Z}_{k}^{R,i}$ for all $k\geq 2$ and $i=1,2$. These sets will be found from the  {\em echograph}  ${\cal E}_K$, defined in terms of ${\cal T}_K$ to be    
$$\displaystyle{{\cal E}_K:=\{ x-t\nu _0(x)/2: (x,x,t)\in {\cal T}_K\} \subset E^2},$$ where $\nu _0:S_0\rightarrow S^1$ denotes the unit inward normal along $S_0$. Define 
$\pi _{{\cal E}_K}:{\cal E}_K\rightarrow S_0$ by taking $\pi _{{\cal E}_K}(w)$ to be the 
point on $S_0$ nearest $w\in {\cal E}_K$, namely $\pi _{{\cal E}_K}(x-t\nu _0(x)/2)=x$.  
\begin{example}\label{ex11} Let $K_1,K_2$ be the regions bounded by the ellipses given implicitly by
$$\frac{4(x_1+\frac{6}{5})^2}{9}+4(x_2+\frac{13}{10})^2=1,\quad \frac{(x_1-x_2)^2}{8}+\frac{(x_1+x_2-1)^2}{2}=1,$$
and choose $S_0$ to be the circle of radius $4$ and centre $c_0=(0,0)$. Figure \ref{echofig} shows a discrete approximation to ${\cal E}_K$ obtained by sampling reflexive trajectories whose middle reflection points are distributed uniformly along $\partial K$. 
$\endofproof$
\end{example}
\spp
The echograph is a union of $C^\infty$ arcs 
meeting in cusps at endpoints, with all other points of intersection transversal. A continuous locally one-to-one function  
$\sigma :\partial K-\{ z_\infty ^{1},z_\infty ^2\} \rightarrow {\cal E}_K$ is given by
$\sigma (z):=x-t\nu _0(x)$ where $\gamma _z(t)=x\in S_0$. Then $\sigma $ maps 
the known subsets ${\cal Z}_1^1$ and ${\cal Z}_1^2$ of $\partial K$ to the arcs  of ${\cal E}_K$ nearest 
$S_0$ (the innermost red and blue respectively in Figure \ref{echofig}).  
%

\spp
Suppose inductively that, for some $k\geq 2$, $\cup _{j<k}({\cal Z}_j^{L,1}\cup {\cal Z}_j^{L,2})$ is determined. Then the corresponding $C^\infty$ arcs $\sigma ({\cal Z}_j^{L,1}),\sigma ( {\cal Z}_j^{L,2})$ in ${\cal E}_K$ are also determined. We might not know ${\cal Z}_k^{L,i}$, but its image 
$\sigma ({\cal Z}_k^{L,i})$ can be seen in the epigraph, since the neighbouring arc ${\cal Z}_{k-1}^{L,i}$ 
is already determined. There is a diffeomorphism $w\mapsto z$ from  the known $\sigma ({\cal Z}_k^{L,i})$ 
onto the unknown ${\cal Z}_k^{L,i}$ where  
$x=\pi _{{\cal E}_K}(w)\in S_0$ is the intersection of $S_0$  
and $\gamma _z([0,\infty ))$. 

\spp
Given $w\in \sigma ({\cal Z}_k^{L,i})$ let $\gamma _0$ be the reflexive geodesic containing $\gamma _z$.  The travelling-time function ${\cal S}_{\gamma _0}:V_{\gamma _0}\rightarrow \R$ of \S \ref{refsec} is determined from $\sigma ({\cal Z}_k^{L,i})$, and therefore so is the initial direction of $\gamma _0$ at $x$, by Lemma \ref{lem1.5}. In other words the terminal direction of $\gamma _z$ at $x\in S_0$ is determined. 

\spp
Now, by Lemma \ref{backlem} and the inductive hypothesis, 
the intersections of $\gamma _z$ with $\partial K$ occur where $\partial K$ is already determined, except for the undetermined initial point $z$. Therefore, working backwards from $x$ and the terminal direction, 
the points $z_1,z_2,\ldots ,z_p$ of intersection (except for $z$) of $\gamma _z$ with $\partial K$ are determined, and so is the direction from $z_1$ to $z$. So $z$ is determined by $\Vert z-z_1\Vert $, 
which is determined by the length of $\gamma _z$, namely by ${\cal S}_{\gamma _0}(x)$ which is also determined. So all points $z$ in ${\cal Z}_k^{L,i}$ are determined, and similarly all of ${\cal Z}_k^{R,i}$. 
This completes the induction. Then $\partial K$ is determined as the closure of 
$$\cup _{k\geq 1}({\cal Z}_k^{L,1}\cup {\cal Z}_k^{L,2}\cup {\cal Z}_k^{R,1}\cup {\cal Z}_k^{R,2}).$$

\renewcommand{\theequation}{\arabic{section}.\arabic{equation}}

\section{Appendix}
\renewcommand{\theequation}{\arabic{section}.\arabic{equation}}

\noindent
{\bf Proof of Lemma \ref{llem2}:} The argument is similar to that in the proof of Proposition 4.1 in \cite{kn:PS2}.
For convenience we will assume that $S_0$ is centred at $0$.
Fix an arbitrary point $x_0 \in S_0$ and two integers $k, m \geq 1$. 
 Let $\omega_0, \omega_0' \in \sn$, and let $\omega(u)$ ($u \in U\subset \R^{n-1}$)
and $\omega(v)$ ($v \in V\subset \R^{n-1}$) be smooth parametrizations of small neighbourhoods of 
$\sn$ near $\omega_0$ and $\omega'_0$, respectively. Assuming $\omega_0 \neq \omega'_0$, we take $U$ and $V$
sufficiently small so that $\omega(u) \neq \omega(v)$ for all $u \in U$ and $v \in V$.

We will consider pairs of simply reflecting
regular $(x_0,y)$-geodesics $\gamma(u)$ and $\gamma(v)$ in $\Omega_K$ issued from $x_0$ in directions 
$\omega(u)$ and $\omega(v)$, respectively, such that $\gamma(u)$ has exactly $k$ reflection points
$x_1(u), \ldots, x_k(u)$ and $\gamma(v)$ has exactly $m$ reflection points $y_1(v), \ldots,y_m(v)$.
Set $x_0(u) = x_0 = y_0(v)$ and $x_{k+1}(u) = y = y_{m+1}(v)$. Then the travelling times of the
two geodesics are
$$f(u) = \sum_{i=0}^k \|x_i(u) - x_{i+1}(u)\| \quad, \quad g(v) = \sum_{j=0}^{m} \|y_j(v)-y_{j+1}(v)\| .$$

Given an arbitrary $r = 1, \ldots,n$,
let $M (k,m,r)$ be the set of those $(u,v)\in U\times V$ for which there exist
$y\in S_0$ with $y^{(r)} \neq 0$ and simply reflecting regular $(x_0,y)$-geodesics in 
$\Omega_K$, $\gamma(u)$ with $k$ reflection points issued from $x_0$ with direction $\omega(u)$ and 
$\gamma(v)$ with $m$ reflection points issued from $x_0$
with direction $\omega(v)$ and  having equal travelling times.

Lemma \ref{llem2} is an immediate  consequence of the following lemma.

\bs

\noindent
\begin{lemma}\label{llem6} For all $k, m \geq 1$ and $r = 1,\ldots, n$ the set $M(k,m,r)$ is a 
smooth submanifold of $U\times V$ of dimension $n-2$.\end{lemma}

\spp
{\bf Proof.} Let $k, m \geq 1$ and let e.g. $r = n$. For $u \in U$, $v \in V$, let
$\gamma(u)$, $\gamma(v)$, $x_i(u)$, $y_j(v)$, $f(u)$, $g(v)$ be as above.
Set $h(u,v) = f(u) - g(v)$, $h_p(u,v) = x^{(p)}_{k+1}(u) - y^{(p)}_{m+1}(v)$
for $p=1, \ldots, n-1$, and consider the map $H : U\times V \longrightarrow \R^{n}$ defined by
$$H(u,v) = (h(u,v), h_1(u,v), \ldots, h_{n-1}(u,v)) .$$
We will show that $H$ is a submersion on $M(k,m,n) = H^{-1}(0)$.

Let $H(u,v) = 0$ and assume that
\begin{equation}
B \grad h(u,v) + \sum_{p=1}^{n-1} A_p \grad h_p(u,v) = 0
\end{equation}
for some $B, A_1, \ldots, A_{n-1} \in \R$.  Set $A_n = 0$ and consider the vector
$A = (A_1, \ldots, A_n) \in \R^n$. Notice that $H(u,v) = 0$ implies $x_{k+1}(u) = y_{m+1}(v)$.

Next, as in the proof of Lemma \ref{llem5}, considering the vectors
$$e_i(u) = \frac{x_{i+1}(u) - x_i(u)}{\|x_{i+1}(u) - x_i(u)\|} \quad, \quad \te_j(v) = \frac{y_{j+1}(v) - y_j(v)}{\|y_{j+1}(v) - y_j(v)\|} ,$$
and using $\frac{\partial x_0}{\partial u_j}  = 0$, $e_0(u) = \omega(u)$ and $\te_0(v) = \omega'(v)$, we get
\begin{eqnarray*}
\frac{\partial f}{\partial u_s} (u) 
& = & - \left\langle  e_0 , \frac{\partial x_0}{\partial u_j} \right\rangle  - \left\langle e_1 - e_0 , \frac{\partial x_1}{\partial u_j} \right\rangle 
- \ldots  \cr
&   & - \left\langle e_{k} - e_{k-1},\frac{\partial x_{k}}{\partial u_j} \right\rangle +
\left\langle e_{k} ,\frac{\partial x_{k+1}}{\partial u_j} \right\rangle =  \left\langle e_{k}(u) ,\frac{\partial x_{k+1}}{\partial u_s}(u) \right\rangle .
\end{eqnarray*}
In the same way, 
$$\displaystyle \frac{\partial g}{\partial v_s} (u,v)=  
\left\langle \te_{m+1}(v) , \frac{\partial x}{\partial v_s}(v) \right\rangle .$$
Now considering in (5.7) the derivatives with respect to $u_s$ for some $s = 1, \ldots,n-1$, we get
$$B \frac{\partial f}{\partial u_s} (u) + \sum_{p=1}^{n} A_p\frac{\partial x^{(p)}_{k+1}}{\partial u_s} (u)  = 0 ,$$
that is
$$B \left\langle e_{k}(u) ,\frac{\partial x_{k+1}}{\partial u_s}(u) \right\rangle 
+ \left\langle A ,\frac{\partial x_{k+1}}{\partial u_s}(u) \right\rangle = 0 ,$$
so $B e_k(u) + A \perp \frac{\partial x_{k+1}}{\partial u_s}(u)$ for all $s = 1, \ldots,n-1$.
However, since the geodesic $\gamma(u)$ is regular, the vectors $\frac{\partial x_{k+1}}{\partial u_s}(u)$
($s=1, \ldots,n-1$) form a basis for the tangent space to $S_0$ at $x_{k+1}(u)$, and therefore 
the vector $B e_k(u) + A$ must be perpendicular to this tangent space, i.e. it is parallel to
$x_{k+1}(u)$ (which is a normal vector to $S_0$ at $x_{k+1}(u)$). Thus, 
\begin{equation}
B e_k(u) + A = \lambda \, x_{k+1}(u)
\end{equation}
for some $\lambda\in \R$. 

Similarly, considering derivatives with respect to $v_s$ in (5.7) we get
$B \te_m(v) + A = \mu \, y_{m+1}(v)$ for some $\mu\in \R$.  Since $x_{k+1}(u) = y_{m+1}(v)$, we now get 
\begin{equation}
B(e_k(u) - \te_m(v)) = (\lambda-\mu) x_{k+1}(u) .
\end{equation}
Since $\omega(u) \neq \omega(v)$ (this is true for all $u\in U$ and $v \in V$ by the choice of $U$ and $V$), 
the geodesics $\gamma(u)$ and $\gamma(v)$ are different, and then 
$x_{k+1}(u) = y_{m+1}(v)$ implies $e_k(u) \neq \te_m(v)$. Thus, there exists a tangent
vector $w$ to $S_0$ at $x_{k+1}(u)$ (i.e. a vector $\perp x_{k+1}(u)$) such that
$\langle e_k(u),w\rangle \neq \langle \te_m(v), w\rangle$. Using this in (5.9) gives $B = 0$.
Now (5.8) reads $A =  \lambda \, x_{k+1}(u)$. However, $A_n = 0$, while $x^{(n)}_{k+1}(u) \neq 0$
by assumption, so we must have $\lambda = 0$ and therefore $A = 0$.

This proves that $H$ is a submersion at any $(u,v) \in H^{-1}(0)$, and therefore
$M (k,m,n)$ is a submanifold of $U \times V$ of codimension $n$, i.e. of dimension $n-2$.
\endofproof

\bs

\noindent
School of Mathematics and Statistics, University of Western Australia, Crawley 6009 WA, Australia\\
E-mails: lyle.noakes@uwa.edu.au, luchezar.stoyanov@uwa.edu.au

\end{document}